\documentclass[dvips,preprint]{imsart}

\usepackage{amsthm,amsmath,latexsym,amssymb,url,bm,graphicx,graphics,subfigure}
\RequirePackage{natbib}

\startlocaldefs
\endlocaldefs

\begin{document}

\begin{frontmatter}

\title{A Short Note on Gaussian Process Modeling for Large Datasets using Graphics Processing Units}
\runtitle{Gaussian Process Modeling using GPUs}


\author{\fnms{Mark} \snm{Franey}\ead[label=e1]{069330f@acadiau.ca}},
\author{\fnms{Pritam} \snm{Ranjan}\ead[label=e2]{pritam.ranjan@acadiau.ca}}
\and
\author{\fnms{Hugh} \snm{Chipman}\ead[label=e3]{hugh.chipman@acadiau.ca}}
\address{Department of Mathematics and Statistics, Acadia University, Wolfville, NS, Canada \\ \printead{e1,e2,e3}}

\runauthor{Franey, Ranjan and Chipman}

\begin{abstract}
The graphics processing unit (GPU) has emerged as a powerful and cost effective processor for general performance computing. GPUs are capable of an order of magnitude more floating-point operations per second as compared to modern central processing units (CPUs), and thus provide a great deal of promise for computationally intensive statistical applications. Fitting complex statistical models with a large number of parameters and/or for large datasets is often very computationally  expensive. In this study, we focus on Gaussian process (GP) models -- statistical models commonly used for emulating expensive computer simulators.
We demonstrate that the computational cost of implementing GP models can be
significantly reduced by using a CPU+GPU heterogeneous computing system
over an analogous implementation on a traditional computing system with no GPU acceleration.  Our small study suggests that GP models are fertile ground for further implementation on CPU+GPU systems.
\end{abstract}





\end{frontmatter}

\section{Introduction} \label{sec:introduction}

Driven by the need for realistic real time computer graphics applications, \emph{graphics processing units} (GPUs) now offer more computing power than \emph{central processing units} (CPUs).  This is accomplished with a large number of (relatively slow) processors that can simultaneously render complex 3D graphical applications by relying on parallelism. Modern GPUs can be programmed to execute many of the same tasks as a CPU.  For tasks with efficient (primarily floating point) parallel implementations, GPUs have reduced execution time by an order of magnitude or more \citep{heteroComputeReview}. Additionally, GPUs are extremely cost effective; delivering far more floating point operations per second (FLOPS) per dollar than multi-core CPUs. This provides a great deal of promise for general purpose computing on graphics processing units (GPGPU), and in particular for computationally intensive statistical applications (e.g., \cite{cudaKmeans}; \cite{cudaMCMC}).

In this note, we explore the merits of heterogeneous computing
(CPU+GPU) over traditional computing (CPU only) for fitting Gaussian
process (GP) models. GP models are popular supervised learning models
\citep{rasmussen_williams2006}, and are also used for emulating computer
simulators which are too time consuming to run for real-time predictions
\citep{gp1}. Fitting a GP model with $n$ simulator runs requires the
determinant and inverse computation of numerous $n\times n$ spatial
correlation matrices. The computational costs of both the determinant
and inverse are $O(n^3)$, which become prohibitive for moderate to
large $n$. We demonstrate that for moderate to large
datasets (thousands of runs),
the implementation on a heterogeneous system can be more
than a hundred times faster. The tremendous speed-ups were achieved by
leveraging NVIDIA's \emph{compute unified device architecture} (CUDA)
toolkit \citep{cuda2}.

This note is intended as a ``proof of concept'' to demonstrate the tremendous time savings that can be achieved for statistical simulations by using GPUs. In particular we show that fitting GP models on a heterogeneous computing platform leads to significant time savings as compared to our previous CPU-only implementation.


The remainder of this note is organized as follows: Section~\ref{sec:gp}
briefly reviews the basic formulation of the GP model. We also highlight the computationally intensive steps of the model fitting procedure in this section. Implementation details for both platforms (CPU + GPU and CPU-only) are discussed in Section~\ref{sec:gpu}.  Comparisons are made via several simulations in Section~\ref{sec:results}, followed by discussion in Section~\ref{sec:conclusion}.  It is clear from this study that the efficiency, affordability, and pervasiveness of GPUs suggest that they should seriously be considered for intensive statistical computation.

\section{Gaussian Process Model} \label{sec:gp}

GP models are commonly used for regression in machine learning, geostatistics and computer experiments.  Here, we adopt terminology of computer experiments \citep{gp1}.  Denote the $i$-th input and output of the computer simulator by $x_i=(x_{i1},...,x_{id})$, and $y_i = y(x_i)$, respectively. The simulator output, $y(x_i)$, is modeled as
\begin{equation}\label{GASP}
    y(x_i) = \mu + z(x_i);
 \qquad i=1,...,n,
\end{equation}
where $\mu$ is the overall mean, and $z(x_i)$ is a GP with $E(z(x_i))=0$, $Var(z(x_i))= \sigma^2_z$, and Cov$(z(x_i),z(x_j)) = \sigma^2_z R_{ij}$. Although there are several choices for $R$ \citep{santner2003daa}, we use the power exponential correlation family
\begin{equation}\label{corr}
   R_{ij} = \mbox{corr}(z(x_i),z(x_j)) =
   \prod_{k=1}^{d}\exp \left\{-\theta_k|(x_{ik}-x_{jk})|^{p}\right\},
   \quad \mbox{ for all }  \quad i,j,
\end{equation}
where $\theta = (\theta_1,...,\theta_d)$ is the vector of hyper-parameters,
and $p \in (0,2]$ is the smoothness parameter.  The squared
exponential $(p=2)$ correlation function is very popular and has good
theoretical properties, however, the power exponential correlation function with
$p<2$ is more stable in terms of near-singularity of $R$ (especially if
the design points are very close, which can occur with large $n$ in small
$d$ - dimensional space). We used $p = 1.95$ for all illustrations in this
note \citep{Kaufman}. Alternatively, a suitably chosen nugget with Gaussian correlation can be used \citep{ranjanNugget} to avoid near-singularity.
One could also use another correlation structure like Mat\'ern \citep{stein1999,santner2003daa}, which is considered to be numerically more stable than power exponential correlation.

We implemented the maximum likelihood approach for fitting the GP model, which requires the estimation of the parameter vector $(\theta_1, ..., \theta_d; \mu, \sigma^2_z)$. The closed form estimators of $\mu$ and $\sigma^2_z$ are given by
\begin{equation}\label{muhat}
   \hat{\mu}(\theta) = {({\bf1_n}'R^{-1}{\bf1_n})}^{-1}({\bf1_n}'R^{-1}Y)
   \ \text{and} \   \hat{\sigma}^2_z(\theta) =
   \frac{(Y-{\bf1_n}\hat{\mu}(\theta))'R^{-1}(Y-{\bf1_n}\hat{\mu}(\theta))}{n},
\end{equation}
which are used to obtain the profile log-likelihood
\begin{equation}\label{profile-likelihood}
  -2\log L_{\theta} \propto \log(|R|) + n \log[(Y-{\bf1_n}\hat{\mu}(\theta))' R^{-1}(Y-{\bf1_n}\hat{\mu}(\theta))],
\end{equation}
for estimating the hyper-parameters $\theta = (\theta_1, ..., \theta_d)$, where $|R|$ denotes the determinant of $R$.

The evaluation of $|R|$ and $R^{-1}(Y-{\bf1_n}\hat{\mu}(\theta))$
for different values of $\theta$ is computationally expensive. One
can save some computation time and increase numerical stability by
using matrix factorization of $R$ (e.g., using LU, QR or Cholesky
decomposition) along with solving the linear systems via back-solves
to obtain $R^{-1}(Y-{\bf1_n}\hat{\mu}(\theta))$. From an implementation
viewpoint, the log-likelihood (\ref{profile-likelihood}) consists of
three computationally expensive components: (i) the evaluation of $R$,
(ii) the factorization of $R$ and (iii) solving the linear systems
via back-solves. The computational cost for the matrix factorization
and back-solves are $O(n^3)$ and $O(n^2)$ respectively. Thus, the
log-likelihood function evaluation is computationally expensive,
specifically for large $n$.

\begin{figure}[h!]\centering
\subfigure[$-2\log L_{\theta}$ surface for $(\theta_1, \theta_2) \in (0, 12)^2$]{\label{fig:old_eg2da_log_lik} \includegraphics[width=3.1in]{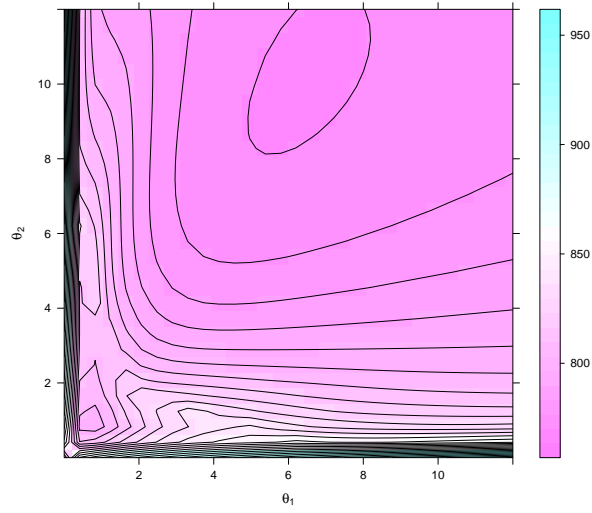}}\\
\subfigure[$-2\log L_{\theta}$ surface zoomed in near $(\theta_1,\theta_2)=0$]{\label{fig:old_eg2db_log_lik} \includegraphics[width=3.1in]{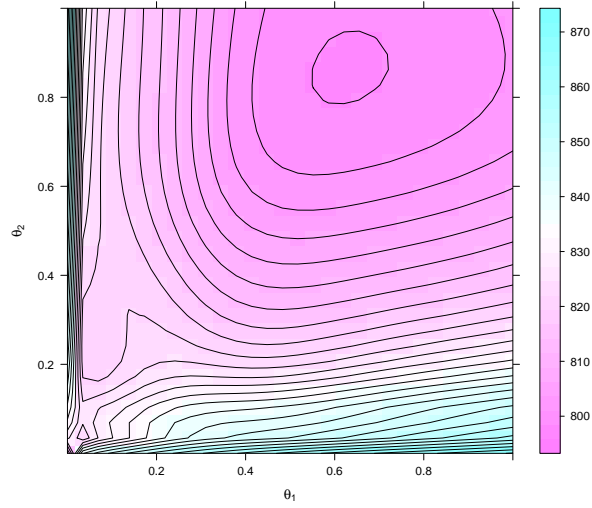}}
\caption{Plots of $-2\log L_{\theta}$ for the GoldPrice function (Example~1) with $n=30$ points chosen using a random maximin LHD.}\label{fig:old_eg2d_log_lik}
\end{figure}

The log-likelihood surface is sometimes
bumpy near $\theta=0$ and often has multiple local optima
\citep{Yuan,Schirru,lawrence,petelin}.  This makes maximizing the likelihood
challenging. For instance, Figure~\ref{fig:old_eg2d_log_lik}
presents contours of $-2\log L_{\theta}$ for a $30$-point, two-dimesional
data set,
where the inputs are generated using a maximin Latin hypercube design
(LHD) in $[0, 1]^2$, and the computer simulator outputs are obtained
by evaluating the two-dimensional GoldPrice function (see Example~1 for
more details on this test simulator).

The log-likelihood surface in Figure~\ref{fig:old_eg2d_log_lik} contains
multiple local minima near $\theta=0$. Though not very clear from
Figure~\ref{fig:old_eg2db_log_lik}, more local minima are found by
enlarging the region near $\theta =0$. A good optimization algorithm that
guarantees the global optimum of such an objective function would require
numerous function evaluations. Evolutionary algorithms like genetic
algorithm \citep{ranjan,ranjanNugget} and particle swarm optimization
\citep{petelin} are commonly used for optimizing the log-likelihood of the
GP models. A multi-start gradient based optimization approach might be
faster but would require careful selection of the starting points to
achieve the global minimum of $-2\log L_{\theta}$.  Once a sufficiently
large number of starting points are chosen, a gradient based optimization is also
likely to involve a large number of evaluations of the likelihood.

The large number of likelihood evaluations, no matter what approach is
taken to search for the MLEs, means that efficient ways to evaluate the
likelihood will be very important.

\section{Implementation of GP on CPUs and GPUs} \label{sec:gpu}

In this section, we discuss the implementations of GP fitting algorithms
in the two computing environments. As in \cite{ranjan,ranjanNugget}, we
use a genetic algorithm (GA) for minimizing $-2\log L_{\theta}$. Other
optimizers might use different strategies, but all approaches require
numerous evaluation of (\ref{profile-likelihood}).

Our implementation of computing $-2\log L_{\theta}$ requires one evaluation of $R$, one Cholesky decomposition of $R$ and seven back-solves involving the Cholesky factors. The version of GA implemented here for optimizing the log-likelihood uses $2000$ evaluations of $-2\log L_{\theta}$. Thus, the GP model fitting procedure requires $2000$ evaluations of $R$, $2000$ Cholesky decompositions and $14000$ back-solves. Furthermore, the GP model predictions on a set of $N$ test points are obtained using the estimated parameters with one evaluation of $R$, one Cholesky decomposition and $2N + 6$ back-solves. Thus, fitting a GP model and obtaining model predictions are computationally expensive, and this paper advocates the use of GPUs to reduce the computational burden.

The heterogeneous (CPU+GPU) workstation for this study includes two
moderately high-end consumer grade components each costing approximately
\$200: a  Intel Core i5 750 CPU and a NVIDIA GTX 260 (core 216) GPU. The
Intel CPU is capable of 4 flops per clock cycle $\times$ 4 cores $\times$
2.66 gigaHertz per core = 42.58 gigaFlops per cycle \citep{intelcpu}.  The
NVIDIA GPU is capable of 3 flops per clock cycle $\times$ 216 cores
$\times$ 1.547 gigaHertz per core = 1002 gigaFlops per cycle
\citep{nvidiagpu}, roughly 24 times more than the Intel Core i5 processor.
The workstation used for comparison is a dual-socket quad-core AMD Opteron
2352 with 32 GB of RAM running Matlab version R2009b. This hardware is
slightly older, but is workstation class instead of consumer class, has
more RAM and twice as many CPU cores.

\textbf{CPU implementation}. Most personal computers available now have
multi-core processors. For example, the somewhat high-end consumer-grade
AMD FX-8120 has 8 cores.  We used a
dual-socket quad core AMD Opteron 2352 based workstation, giving 8 cores in
total. The CPU
implementation was written in Matlab, making use of several built-in
functions (e.g., matrix factorization routines, backslash operator for
back-solving linear systems) that are linked to compiled high performance
subroutines written in Fortran or C. The Cholesky decomposition was used
for computing both the determinant and the inverse of $R$ (via solving
linear systems).  The Matlab (CPU) code did not contain specialized
parallelization commands.  However, the Matlab session used all 8 cores
through its intermal math routines.

\textbf{CPU+GPU implementation}. The most popular GPGPU platform is
that of a Intel or AMD x86 CPU and a Nvidia GPU.  This is the platform
that was adopted for this study - an Intel core i5 750 quad core CPU and
Nvidia GTX 260 GPU. We used the Nvidia GPGPU software development kit
(CUDA 2.3) to implement the Gaussian process model on this system.  As part
of that implementation we
rewrote our Matlab code in the C for CUDA programming language
to enable general purpose computation on the GPU. The free libraries
\emph{CUDA basic linear algebra subroutines} (CUBLAS) \citep{cublas} and
CULATools \citep{cula}
were used
for executing the most computationally expensive steps of the GP model on
the GPU.
CULATools is a proprietary library with a free version that
provides popular linear algebra operations for NVIDIA GPUs.
Because the Cholesky decomposition routine was not part of the
free CULATools library, the free (but closed source) LU decomposition
(\texttt{culaDeviceSgetrf}) was used instead.  Though we are not certain due
to the proprietary nature of CULATools, we believe the GPU LU decomposition
is parallelized in a block-based manner similar to the freely available
SCALAPACK version for CPUs \citep{choi1992scalapack}.

We explicitly used the GPU for evaluation of correlation matrix $R$ using a
custom GPU kernel (C for CUDA function), LU decomposition of $R$ (CULATOOLS
library), all linear algebraic operations including the back-solves (CUBLAS
library), and a few other custom C for CUDA functions in order to minimize
the transfer of data to and from the GPU. The
computation of $R$ requires $d\cdot n(n-1)/2$ evaluations of $-\theta_k|(x_{ik} - x_{jk})|^{1.95}$ all of which are mutually independent and can be executed in parallel.

Note that no task parallelization (e.g., multiple evaluation of the
likelihood at different parameter values)
was performed in this implementation. Evaluating the likelihood for numerous parameters simultaneously seems a natural approach to parallelizing a GP model, however this type of parallelism was not possible because only one kernel could have been executed at a time. If we had access to a more advanced GPU (e.g., GF100 series or newer), executing some of these computationally intensive steps in a task parallel manner (using \emph{concurrent kernel execution}) could potentially lead to even more time savings.

To summarize, we used the GPU for all linear algebra operations and building the
correlation matrix $R$. The CPU was used for data input/output and
process control.
Additional execution configurations could have been tuned to optimize
the performance of GPUs \citep{cudaMBA}. However the aim of this study
is not to achieve the highest performance possible for these models and
procedures; but rather to explore the use of a GPU as a co-processor that
can lead to significant time savings at a reasonable cost.

\section{Examples}\label{sec:results} \label{sec:gpFitResults}

In this section, we use simulated examples to compare the results
obtained from the two implementations of the GP model. Throughout this
note, the
CPU implementation of the methodology in Matlab is referred
to as \emph{CPU implementation} and the mixed system (CPU + GPU)
implementation (in C based language called CUDA) is referred to as
\emph{CUDA implementation}. Our CUDA implementation enforces single
precision floating-point arithmetic and uses LU instead of Cholesky
decomposition of the correlation matrices $R$.

For fair comparison of the results, both implementations used exactly the
same input data and the number of likelihood evaluations in the
optimization procedure. The design points, $\bm{x}=(x_1,...,x_n)$, were
generated using a maximin Latin-hypercube scheme. For efficient distribution of the data to the processors in the CUDA implementation, we used powers of $2$ for $n$, however, other values of $n$ can also be considered in our implementations (e.g., we used $n=4064$ in Example~2). Simulations with $n > 4070$ were not considered here because of a restriction within the CUBLAS library (version 2.3). In the most recent CUBLAS release (3.0) this restriction has been relaxed, allowing for much larger matrices to be considered in future applications.

As pointed out in Sections~\ref{sec:gp} and \ref{sec:gpu}, the challenge of effectively
maximizing the likelihood for a GP model leads to algorithms that
repeatedly evaluate the likelihood.  In the examples considered in this
section, a genetic algorithm with 20 generations and a population of 100
solutions leading to 2,000 likelihood evaluations, was used.

\textbf{Example 1.} Suppose the emulator outputs are generated from the
$\log$ of the 2-dimensional Goldstein-Price function \citep{hartmanSix}.
Designs of different sizes, $n=2^k$ for $k=4, 5,...,9$, were used to fit
the GP model. The reason for not using larger designs in this simulation
study is discussed in Section~5. Table~\ref{tab:gpFitGoldpriceResults}
shows the parameter estimates, optimized value of $-2\log(L_{\theta})$ as
in (\ref{profile-likelihood}), sum of squared prediction error ($SSPE =
\sum_{i=1}^{1000} (y_i-\hat{y}_i)^2$), and total runtime of the code (i.e.,
fitting the GP model and computing SSPE). The SSPE is calculated over a
fixed test set, which  is a maximin LHD chosen independent to the design used for model fitting.
Both CPU and CUDA implementations used the same training and test datasets. A slight randomness (or variation) in these performance measures is introduced due to the stochasticity of the genetic algorithm used for optimizing the likelihood. To reduce the variation, the results in Table~\ref{tab:gpFitGoldpriceResults} are averaged over ten simulations. For each replication of the experiment a different design (training dataset for fitting the GP model) was chosen.

\begin{table}[!h]\centering
\caption{Simulation results for Goldstein-Price function, averaged over 10 replications.}
\begin{tabular}{|c|}
\hline
CPU implementation\\
\begin{tabular}{rrrrrr}
 \hline
  $n$   & Time(sec) & $-2log L_{\theta}$ & $\hat{\mu}$ & $\hat{\sigma}^2_z$ & SSPE \\ \hline
  16  & 0.91 	& 55.95 	 & 9.80 & 7.09 & 1158.89 \\
  32  & 3.30 	& 130.98     & 9.36 & 5.19 & 860.47 \\
  64  & 12.90 	& 286.46 	 & 9.47 & 5.31 & 219.73 \\
  128 & 51.45 	& 541.83 	 & 9.58 & 4.77 & 131.87 \\
  256 & 206.18  & 1186.40    & 9.31 & 3.17 & 28.84 \\
  512 & 911.93  & 2287.88    & 9.41 & 4.91 & 11.91 \\
%
\end{tabular}\\
\hline
\hline
CUDA implementation\\
\begin{tabular}{rrrrrr}
  \hline
   $n$ & Time(sec) & $-2log L_{\theta}$ & $\hat{\mu}$ & $\hat{\sigma}^2_z$ & SSPE \\  \hline
  16  & 2.26 & 56.85 	 & 9.83 & 7.19 & 1188.21 \\
  32  & 4.94 & 130.48    & 9.47 & 5.25 & 831.98 \\
  64  & 5.50 & 285.47 	 & 9.43 & 4.61 & 225.83 \\
  128  & 7.25 & 536.32   & 9.54 & 3.61 & 134.15 \\
  256  & 9.88 & 1207.87  & 9.54 & 4.73 & 40.56 \\
  512  & 20.75 & 2321.21 & 9.59 & 4.79 & 14.38 \\
  \end{tabular}\\
 \hline
\end{tabular}
\label{tab:gpFitGoldpriceResults}
\end{table}

It is clear from Table~\ref{tab:gpFitGoldpriceResults} that the optimized
value of the profile log-likelihood ($-2\log L_{\theta}$), $\hat{\mu}$,
$\hat{\sigma}^2_z$ and SSPE from the two implementations are comparable
within each sample size.  The discrepancy between the
results of these two implementations may be partially explained by
the fact that single precision floating point operations were used in
CUDA as compared to double precision calculations in Matlab.  See
Section~\ref{sec:conclusion} for further discussion.


While the discrepancies in Table~\ref{tab:gpFitGoldpriceResults} suggest
that single-precision arithmetic on the GPU may not always yield exact
optima, we stress that the most difficult part of the likelihood
optimization is finding the right neighbourhood of a good optimum.  Short
double-precision runs on the CPU may be used to refine the CUDA results.

The column ``Time(sec)" in Table~\ref{tab:gpFitGoldpriceResults} shows that
as $n$ increases the CUDA implementation significantly outperforms the
CPU implementation.  For $n=512$, on average CPU implementation takes more than $15$ minutes but CUDA implementation requires roughly $21$ seconds.

\textbf{Example 2.} Consider the 6-dimensional Hartman function \citep{hartmanSix} for generating the simulator outputs. Since the input space is six dimensional, we considered relatively large designs with $n=2^6, 2^8, 2^{10}$ and $4064$ ($n>4070$ is not allowed due to CUBLAS restrictions) to fit the GP model, and the SSPE was calculated over 1000 points chosen using a maximin LHD. Table~\ref{tab:gpFitHartmanResults} summarizes the simulation results.

\begin{table}[!h]\centering
\caption{Simulation results for Hartman function, averaged over 10 replications (except *, one replication)}
\begin{tabular}{|c|}
\hline
CPU implementation\\
\begin{tabular}{rrrrrrrrr}
  \hline
  $n$  & Time(s) & $-2log L_{\theta}$ 	&  $\hat{\mu}$ & $\hat{\sigma}^2_z$  & SSPE \\
  \hline
    64 &  32.32 & 125.94 & 0.1771 & 0.1403 & 77.4160 \\
  256 &  514.43 & 610.25 & 0.1105 & 0.1164 & 27.4311 \\
  1024 &  13325.86 & 2491.97 & 0.0609 & 0.0970 & 5.6504 \\
4064 & *161925.05 & *8044.80 & *0.0485 & *0.0824 & *0.5320 \\
\end{tabular}\\
\hline
\hline
CUDA implementation\\
\begin{tabular}{rrrrrrrrr}
  \hline
   $n$  & Time(s) & $-2log L_{\theta}$ 		&  $\hat{\mu}$ & $\hat{\sigma}^2_z$  & SSPE \\
  \hline
    64 &  9.45 & 103.70 & 0.1238 & 0.2989 & 91.7860 \\
  256 & 16.58 & 547.10 & 0.1397 & 0.1746 & 31.4641 \\
  1024 &  96.19 & 2665.58 & 0.1192 & 0.1390 & 4.3850 \\
  4064 &  1059.71 & 8698.28 & 0.0803 & 0.0700 & 0.5314 \\
\end{tabular} \\
 \hline
\end{tabular}
\label{tab:gpFitHartmanResults}
\end{table}

Similar to Example~1, the results are averaged over 10 simulations. One
exception is CPU with $n=4064$: a single run took almost 45 hours, and thus we did not perform 10 simulations.
Table~\ref{tab:gpFitHartmanResults} shows that the CUDA implementation achieves computation time savings of up to a factor of 150 ($n=4064$).



\section{Discussion} \label{sec:conclusion}

This note illustrates that GPU parallelization can significantly reduce the computational burden in fitting GP models.  The largest improvements are evident in the most difficult problems (e.g., from 45 hours to 18 minutes in Example~2, $n=4064$ case).  These dramatic savings suggest that more complete implementations of GP models on GPU platforms deserve further study.

Although this article focuses on speedups due to hardware, there
are other approaches for estimation of GP models with large sample
sizes. For instance, ad-hoc methods have been proposed which  modify
the correlation function and give sparse correlation matrices,
facilitating faster computation of the inverses and determinants
(e.g., \cite{furreretal2006}, \cite{stein2008}). However, it seems
quite plausible that efficiencies realized by a GPU+CPU approach would also
benefit these sparse approximations.

A number of design choices in the experiments merit commentary.
In Example 1, the reason for not considering $n=2^{10}$ or larger is that the correlation matrices are near-singular.  Although choice of the exponent in the correlation matrix $R$, set at $p=1.95$, substantially reduces the chance of getting near-singular correlation matrices, this is still a problem for large $n$ in the two-dimensional problem. We did not want to further lower the value of $p$, as it would affect the smoothness of the GP fit. Alternatively, a nugget based approach \citep{ranjanNugget} can be used to implement the GP model with large datasets. It is expected that the runtime of the model fitting procedure would not change significantly with $p$, except when the correlation matrix becomes sparse.

As noted in Section~\ref{sec:results}, MLEs obtained by the
CPU+GPU implementation differed somewhat from those found by the
CPU implementation.  A significant reason for this is the use of
single-precision arithmetic in the GPU code.  These estimates could
be refined by a short run of double-precision CPU code.  Since this
refinement might involve 10 or 20 likelihood evaluations, a small fraction
of the 2000 necesary to find a good solution, we have not included this
effect in our studies.

There are a number of specific functions for high-level software
packages like Matlab and R that leverage GPUs and can be used to improve
existing codes. For instance, our CPU implementation can be improved by using Jacket \citep{jacket} and GPUlib \citep{gpulib}. This approach was not taken for this study as GPUs were not available in our cluster, and Matlab was not available on our heterogenous computing system. Similarly, GPUtools is a library for GPU accelerated routines in R \citep{gputools}.

Though not discussed here, it is possible to make use of more than one CPU and as many as three or four GPUs in one application/system, if there are available GPU slots. In fact, cutting edge research is currently being done using multiple GPUs coupled by an extremely fast interconnect (Infiniband) \citep{infinibandGPU}.  Even more processing power can be achieved by creating a cluster of CPU+GPU systems coupled over a network, like the second and fourth fastest supercomputers in the world (as of June 2011), the Chinese Tianhe-1A and Nebulae \citep{top500}.  Developing algorithms to execute in parallel across numerous processors is not a task that is frequently undertaken by statisticians. However, parallel computing, GPU co-processing, and heterogeneous computing techniques can become popular in the near future for computationally intensive statistical computing applications because of the promise of heterogeneous computing and likely changes to CPU architectures \citep{heteroComputeReview}.


\bibliographystyle{agsm}
\bibliography{hello}


\end{document}